\documentclass[twocolumn,showpacs,amsmath,amssymb,prb,superscriptaddress]{revtex4}

\usepackage{graphicx}
\usepackage{color}

\usepackage{hyperref}

\begin{document}

\title{Magnetic order in $\alpha$-RuCl$_3$: a honeycomb lattice quantum magnet with strong spin-orbit coupling}

\author{J. A. Sears}
\author{M.~Songvilay}
\author{K. W. Plumb}
\author{J. P. Clancy}
\affiliation{Department of
Physics, University of Toronto, 60 St.~George St., Toronto, Ontario,
M5S 1A7, Canada}
\author{Y. Qiu}
\author{Y. Zhao}
\affiliation{NIST Center for Neutron Research, National Institute of Standards and Technology, Gaithersburg, Maryland, 20899, USA}
\affiliation{Department of Materials Science and Engineering, University of Maryland, College Park, Maryland 20742, USA}
\author{D. Parshall}
\affiliation{NIST Center for Neutron Research, National Institute of Standards and Technology, Gaithersburg, Maryland, 20899, USA}
\author{Young-June~Kim}
\email{yjkim@physics.utoronto.ca} \affiliation{Department of
Physics, University of Toronto, 60 St.~George St., Toronto, Ontario,
M5S 1A7, Canada}

\date{\today}

\begin{abstract}
We report magnetic and thermodynamic properties of single crystal $\alpha$-RuCl$_3$, in which the Ru$^{3+}$ ($4d^5$) ion is in its low spin state and forms a honeycomb lattice. Two features are observed in both magnetic susceptibility and specific heat data; a sharp peak at 7~K and a broad hump near 10-15K. In addition, we observe a metamagnetic transition between 5~T and 10~T. Our neutron diffraction study of single crystal samples confirms that the low temperature peak in the specific heat is associated with a magnetic order with unit cell doubling along the honeycomb (100) direction, which is consistent with zigzag order, one of the types of magnetic order predicted within the framework of the Kitaev-Heisenberg model.
\end{abstract}


\maketitle

Physics driven by spin-orbit coupling (SOC) is drawing much attention these days. \cite{Witczak2014,Kitaev2006,Wan2011,Okamoto2007,Lawler2008,Shitade2009,Chaloupka2010,Pesin2010,Witczak2011}
In particular, the Kitaev model, a spin-1/2 model on a honeycomb lattice with bond-dependent spin interactions, has captured the interest of both the quantum computing and condensed matter communities. \cite{Kitaev2006,Jackeli2009,Chaloupka2010,Reuther2011,Trousselet2011,Kimchi2011,Subhro2012,Singh2012,Price2012,Price2013,Rau2014,Yamaji2014} The ground state of this model is an exactly solvable quantum spin liquid, and supports gapless excitation of Majorana fermions \cite{Kitaev2006}. Unlike spin liquids found in geometrically frustrated quantum magnets, the Kitaev spin liquid arises from the bond-dependent interactions that frustrate spin configuration on a single site. Since such bond-dependent interactions naturally exist in materials with strong SOC, iridates and other 5d transition metal compounds have been intensively investigated.\cite{Singh2010,Singh2012,Choi2012,Ye2012,Liu2012,Comin2012,Gretarsson2013,Kimchi2014,Modic2014} In particular, honeycomb lattice iridates $\rm A_2IrO_3$ (A=Na or Li) have been extensively scrutinized. In real materials, bond-dependent symmetric off-diagonal exchange ($\Gamma$) as well as the isotropic Heisenberg interaction ($J$) are invariably present in addition to the Kitaev term ($K$), and a realistic spin Hamiltonian for the Kitaev-Heisenberg (KH) model requires all three $J$-$K$-$\Gamma$ terms \cite{Jackeli2009,Chaloupka2010,Rau2014,Katukuri2014}. We note that alternative interpretations, such as the quasi-molecular orbital (QMO) model, have also been proposed to describe the iridates. \cite{Mazin2012}

Recently it was pointed out that $\alpha$-RuCl$_3$ is another model magnetic system on a honeycomb lattice, in which the KH model might be applicable.\cite{Plumb2014b}  The magnetic moment in this material arises from the Ru$^{3+}$ ($4d^5$) ion at the center of an RuCl$_6$ octahedron. The Cl-Ru-Cl angles and the Ru-Cl bond lengths seem to suggest that the octahedron is close to an ideal one, which means that the additional (trigonal, tetragonal, etc.) crystal fields are negligible compared to the RuCl$_6$ octahedral crystal field. As a result, the SOC plays an important role in $\alpha$-RuCl$_3$ even though the bare SOC value is smaller than that in iridates, and the magnetic state of the Ru$^{3+}$ ion is described by the same $J_\mathrm{eff}=1/2$ state as in the  iridate materials. The Ru-Cl layers are stacked along the c-direction to form a CrCl$_3$ type structure $P3_112$ (\#151) \cite{Stroganov1957}. The layers are only weakly bonded with van der Waals interaction, resulting in a very micaceous material.

The electronic properties of this material have been studied at length over the years. \cite{Binotto1971,Rojas1983,Guizzetti1979,Pollini1994,Pollini1996}  While early transport studies described $\alpha$-RuCl$_3$ as a small gap semiconductor, later spectroscopic studies seem to favor a Mott insulator description.\cite{Pollini1996} These conflicting viewpoints were resolved in a recent study, in which it was pointed out that the Mott insulating behavior of $\alpha$-RuCl$_3$ arises from the combined effect of band narrowing due to SOC and moderate size electron correlation.\cite{Plumb2014b} Magnetic properties of $\alpha$-RuCl$_3$ were first reported by Fletcher et al., who found a sharp cusp around 13-15~K in their magnetic susceptibility data on a powder sample, which was attributed to a magnetic phase transition.\cite{Fletcher1967} Similar results were obtained in later studies by Kobayashi et al. \cite{Kobayashi1992} However, no detailed information about the nature of magnetic ground state of this compound has been reported.

In this article, we report a comprehensive examination of the magnetic properties of $\alpha$-RuCl$_3$ single crystals. We found a magnetic phase transition at 7~K in our neutron diffraction, magnetic susceptibility, and specific heat measurements. The magnetic order is described by a doubling of the unit cell along the hexagonal (100) direction, which is consistent with the zigzag ground state of a honeycomb magnet. However, the estimated ordered moment is small, and the magnetic order is short-ranged along the direction perpendicular to the honeycomb plane. When magnetic field is applied along the c-direction, we also found a metamagnetic transition due to possible spin-flop. In addition, a prominent and broad feature is observed in both susceptibility and specific heat data around 10-15K, just above the magnetic transition temperature, suggestive of a two-step phase transition. These observations, taken together, suggest that the magnetic ground state of $\alpha$-RuCl$_3$ is quite unusual, and support the claim that this material is in the proximity of a quantum spin liquid phase as proposed in a recent Raman scattering study \cite{Sandilands2014}.

Single crystal samples were prepared by vacuum sublimation from commercial $\rm RuCl_3$ powder. A typical crystal is a thin hexagonal plate about $\sim 1$~mm$^2$ in area. An example is shown in the inset of Fig. 1(a). Magnetic susceptibility measurements were carried out using a Magnetic Property Measurement System (MPMS). Magnetization curves were obtained at different temperatures using a Physical Property Measurement System (PPMS) with fields up to 14 Tesla. Specific heat measurements were also carried out using the PPMS in zero applied field. Neutron diffraction measurements were carried out using the Multi-Axis Crystal Spectrometer (MACS) and the BT-7 triple axis spectrometer at the NIST Center for Neutron Research (NCNR). The measurements on MACS were conducted using a collection of 35 single crystal samples mounted together with a mosaic width of about 10 degrees and total mass of 62 mg.  Incident neutron energy was 5 meV, and the sample was mounted in the (H0L) plane. The BT-7 data was collected using a similar crystal array of 60 crystals, with a mass of 70 mg. The incident neutron energy was 14.7 meV, and measurements were conducted in both the (H0L) and (HHL) planes. Throughout this paper, we use the hexagonal notation of $a=5.96$~\AA\ and $c=17.2$~\AA\ \cite{Stroganov1957}.

\begin{figure}
\includegraphics[width=0.9\columnwidth]{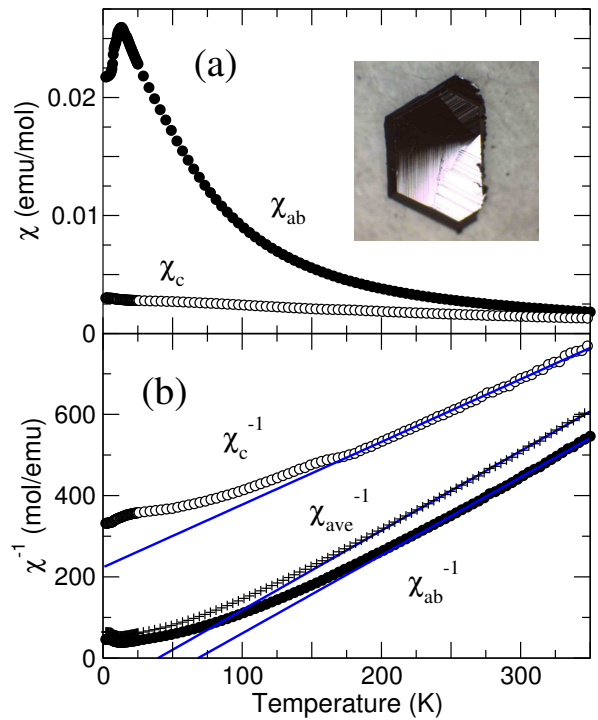}
\caption{(Color online) (a) Temperature dependence of magnetic susceptiibilites with a field of 0.5 T applied parallel to the c-axis and within the ab-plane. Inset shows a photograph of a typical single crystal sample used in our studies. (b) Inverse susceptibility in both directions. Also plotted is the inverse of powder average susceptibility $\chi_{ave} \equiv (2 \chi_{ab} + \chi_c)/3$.
} \label{fig1}
\end{figure}

The temperature dependence of the magnetic susceptibility measured with a 0.5~Tesla field is shown in Fig.\ref{fig1}(a). Here we use the notation $\chi_{c}$ to denote susceptibility measured with field applied perpendicular to the honeycomb plane, and $\chi_{ab}$ for susceptibility measured with in-plane field. The \(\chi_{ab}\) data exhibit a peak around 15 K, in agreement with earlier reports on powder samples \cite{Fletcher1967,Kobayashi1992}. The susceptibility is highly anisotropic; \(\chi_{ab}\) is almost an order of magnitude larger than \(\chi_{c}\) at low temperatures. The Curie-Weiss temperatures also differ significantly in the two directions. In Fig.~\ref{fig1}(b), the inverse susceptibility data are fitted with Curie-Weiss behavior above 200~K. The Curie-Weiss temperatures are $\Theta_{c} \approx -145$~K  and $\Theta_{ab} \approx 68$~K. The effective paramagnetic moments inferred from the Curie constant fit of the susceptibility are \(\mu_{eff} \approx 2.0 \mu_{B}\) and \(\mu_{eff} \approx 2.3 \mu_{B}\) for \(\chi_{ab}\) and \(\chi_{c}\) respectively. These values for paramagnetic moments are consistent with earlier reports, and are larger than the spin-only value of 1.73\(\mu_{B}\) for the low-spin state (S=1/2) for Ru$^{3+}$, which probably indicates a significant contribution from the orbital moment. Although the Curie-Weiss temperatures obtained in our study are different from the values reported earlier, when we fit the powder average ($\chi_{ave} \equiv (2 \chi_{ab} + \chi_c)/3$), we obtain Curie-Weiss temperature of about 40~K, more in line with earlier studies \cite{Fletcher1967,Kobayashi1992}. The observed anisotropy of Curie-Weiss temperature has an interesting implication in view of the $J$-$K$-$\Gamma$ model. According to the high temperature expansion formula introduced in Ref.~\cite{Rau2014}, the Curie-Weiss temperature anisotropy satisfies $(\Theta_c-\Theta_{ab})/(\Theta_c+2\Theta_{ab})=\Gamma/(3J+K)$. Since we find $\Theta_c \approx - 2\Theta_{ab}$, and we assume that $\Gamma$ is not infinitely large, we can estimate that $J \sim -K/3$ in this compound. This is quite different from $\rm Na_2IrO_3$, for which $\Gamma/(3J+K) \sim -0.3$ \cite{Singh2010,Rau2014}. We also note that the susceptibility anisotropy of $\alpha$-RuCl$_3$ is opposite to that of $\rm Na_2IrO_3$; that is, $\chi_{ab} < \chi_{c}$ in $\rm Na_2IrO_3$.

\begin{figure}
\includegraphics[width=0.9\columnwidth]{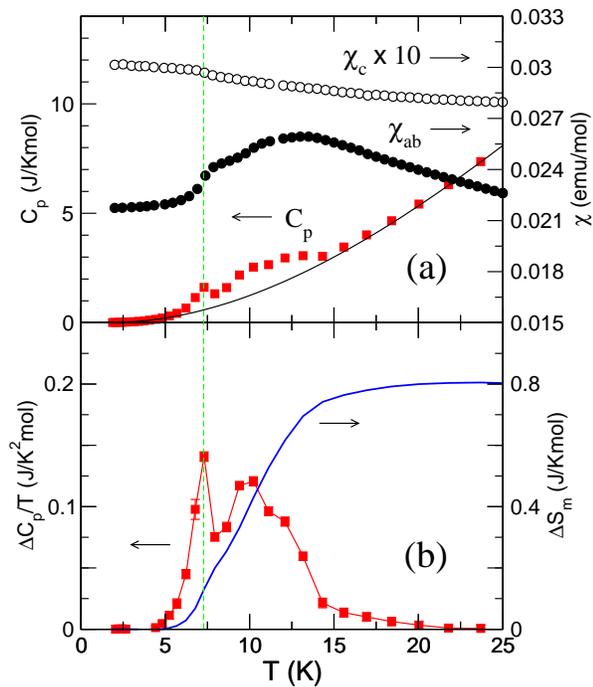}
\caption{(Color online) (a) Detailed view of low temperature region of the parallel and perpendicular susceptibilities shown in Fig.~1(a). $\chi_{c}$ values are multiplied by 10 to fit on the same scale. Also shown is the temperature dependence of heat capacity. (b) The non-monotonic part of the heat capacity, obtained by subtracting a smooth polynomial background (i.e., the solid line in panel (a)) from the raw data. The entropy change $\Delta S_m$ was obtained by integration: $\Delta S_m=\int_0^T (\Delta C_p/T) dT$.
} \label{fig1-2}
\end{figure}

In Fig.~\ref{fig1-2}(a), the transition region below 25 K is magnified. There are clearly two features, a sharp suppression of susceptibility below 7~K and a broad peak at higher temperatures around 10-15~K. The 7~K anomaly is also observed in the $\chi_c$ data, which is shown here by multiplying a scale factor of 10. During our studies, we noticed that the size of low temperature $\chi_c$ showed fairly significant variations depending on the crystal growth condition. The anomaly around 150~K observed in $\chi_c^{-1}$ shown in Fig.~\ref{fig1}(b) is also sample dependent. Further investigation to understand this behavior is under way. Nevertheless the magnetic transition temperature around 7~K, and the two-peak profile of $\chi_{ab}$ at low temperatures do not vary across different crystals. To confirm the existence of the two features observed in susceptibility, we have carried out specific heat measurements in this temperature range as shown in Fig.~\ref{fig1-2}(a).

The specific heat data at higher temperatures (20K-50K) were fit to a smooth polynomial function \cite{Cp_note},
and subtracted from the data to show the magnetic contribution ($\Delta C_p$) as shown in Fig.~\ref{fig1-2}(b). The two peak feature, a sharp low temperature peak accompanied by a broad high temperature hump is remarkably similar to the magnetic susceptibility behavior. The transition temperature around 7~K coincides with the temperature at which a peak is observed in $\chi_{ab}$ as illustrated with the vertical dashed line in Fig.~\ref{fig1-2}.

\begin{figure}
\includegraphics[width=\columnwidth]{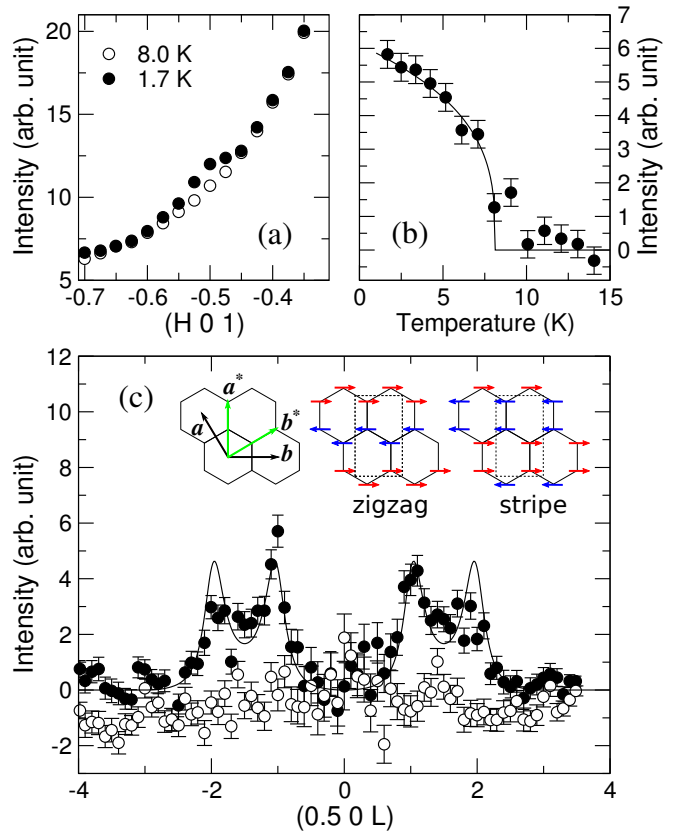}
\caption{(a) Scans across the (-0.5, 0, 1) peak position obtained at two temperatures, 1.7 K and 8K. A peak is observed at 1.7K. A large sloping background is due to the unscattered neutron beam. (b) Temperature dependence of the integrated intensity of the peak shown in panel (a). A background scan obtained at 20K was subtracted from the raw data before the counts were added up. The solid line is a fit to $\sim (T_c-T)^{2 \beta}$ with $\beta=0.2 \pm 0.1$ and $T_c$=8.2(5)~K. (c) $L$-dependence of the magnetic peak at two temperatures. The error bars in the figure represent one standard deviation. The solid line is based on Hendricks-Teller calculation as described in the text. The inset shows the unit vectors in real and reciprocal space, and spin arrangements in the zigzag and stripe ordering pattern.
} \label{fig2}
\end{figure}

To further elucidate the nature of the observed magnetic phase transition, we have carried out neutron diffraction experiments on single crystal samples.  The results are shown in Fig.~\ref{fig2}. A clear peak is observed at the (-0.5, 0, 1) position, which disappears when the sample temperature is raised above 8~K. Measurements carried out in the HHL-type plane 90$^\circ$ away from the observed magnetic peaks did not show peaks at (0.5, 0.5, L) type positions. Detailed temperature dependence of the (-0.5, 0, 1) peak can be found in Fig.~\ref{fig2}(b), in which we plot the integrated intensity as a function of temperature after the background scan at T=20~K was subtracted \cite{dave}. Based on the temperature dependence of this peak, we assign this to be a superlattice peak arising from the magnetic order at the 7K transition. It is difficult to extract precise critical behavior due to poor statistics; however, a rough fit to a power law as shown in the figure suggests that the transition is continuous and the transition temperature is $T_N \approx 8 \pm 1$~K. The peak is very broad along the $L$-direction as shown in Fig.~\ref{fig2}(c). The observed $L$-dependence suggests that the order along the c-direction is only short-ranged due to the prevalence of stacking faults. Similar stacking disorder (or partial order) is found in graphite. In their classic work, Hendricks and Teller considered a model for graphite layer stacking, in which the preferred ABAB... type stacking pattern is randomly mixed with the ABCABC... type, and they showed that the structure factor depends only on the probability ratio $x$ of the two types of stacking \cite{Hendricks+Teller}. In Fig.~\ref{fig2}(c), we plot Eq.~(34) from Ref.~\cite{Hendricks+Teller}. The remarkable agreement with only intensity scaling strongly indicates that a similar type of stacking disorder is present in $\alpha$-RuCl$_3$. The value of the parameter $x=2.7$ to describe our data indicates that the ABCABC type stacking is 3 times more prevalent than the ABAB type stacking, in agreement with the $P3_112$ structure \cite{Stroganov1957}. This is not surprising, since the crystal morphology of $\alpha$-RuCl$_3$ is quite similar to graphite. We would like to note that the limited magnetic correlation along the $c$-direction has a structural origin, since similar $L$-dependence is also observed for {\em structural} Bragg peaks.

Our neutron data demonstrates the existence of a non-trivial magnetic order below 7K in this compound. Ground states of various parameter regimes of the KH model have been extensively investigated in the context of $\rm Na_2IrO_3$ \cite{Singh2010,Singh2012,Choi2012,Ye2012,Rau2014,Liu2011,Chaloupka2010,Chaloupka2013,Kimchi2011,Subhro2012}. Since the honeycomb lattice is a bipartite lattice, a naive expectation for a Heisenberg magnet ($J$ only) would be either ferromagnetic or Neel order. However, in the presence of $K$ and $\Gamma$ terms, it was found that stripe, zigzag, or spiral order can be stabilized depending on the sign and magnitude of the interactions \cite{Chaloupka2010,Chaloupka2013,Rau2014}. The in-plane ordering wavevector of (0.5, 0) suggests that the unit cell is doubled along the hexagonal $a$-direction. We have also studied the (HHL)-type plane 90$^\circ$ away from the observed magnetic peaks and were not able to detect (0.5, 0.5, L) type peaks. These findings are consistent with zigzag order, which has a structure factor of zero for all (0.5, 0.5) type positions (See Fig.~\ref{fig2}(c) inset), but not with stripe order which has a non-zero structure factor.  Zigzag order puts $\alpha$-RuCl$_3$ in the large antiferromagnetic $K$ and ferromagnetic $J$ regime in the $J$-$K$-$\Gamma$ phase diagram, which is consistent with our estimation of $J \sim -K/3$ obtained from magnetic susceptibility anisotropy. \cite{HYK}

Further refinement of magnetic peaks with improved statistics will be necessary to determine moment size and direction. One can nevertheless gain some insight by considering the following points.
1) {\em ordered moment size}: One can obtain a rough estimate of the ordered moment by comparing the intensity of the observed magnetic Bragg peak with that of a structural Bragg peak. We estimate that the ordered moment is at least an order of magnitude smaller than the full paramagnetic moment. Such a small moment size indicates that the order is susceptible to large quantum fluctuations. It should be noted that this is a rough estimate only, as it was based on a structural model that does not account for stacking faults. 2) {\em ordered moment direction}: The behavior of the susceptibility below the ordering temperature, a large drop in $\chi_{ab}$ and virtually constant behavior for $\chi_c$, is consistent with having a sizable component of the ordered moment in the honeycomb plane. However, the finite $T=0$ value of $\chi_{ab}$ as well as the magnetization data discussed below suggest that the magnetic moment component perpendicular to the plane is also non-zero.

\begin{figure}
\includegraphics[width=0.95\columnwidth]{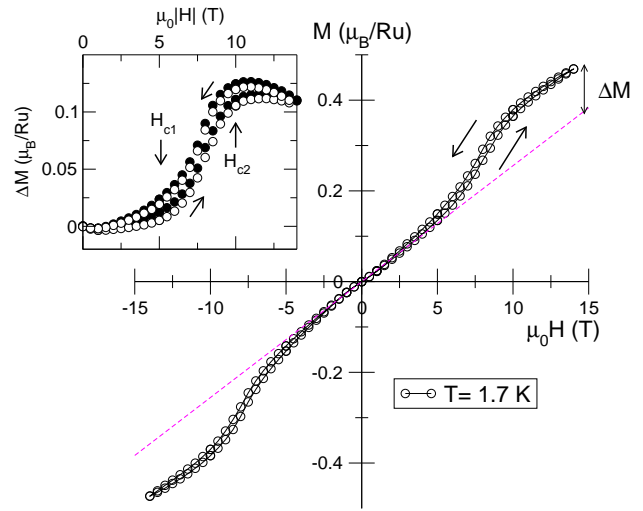}
\caption{Magnetization isotherm at 1.7 K measured with magnetic fields applied perpendicular to the honeycomb lattice. The inset shows $\Delta M$ for both positive (empty circles) and negative (filled circles) field directions, obtained by subtracting linear dependence (dashed line) from the raw data.
} \label{fig3}
\end{figure}

We present our magnetization study as a function of applied field in Fig.~\ref{fig3}. We measured the magnetization isotherms at 1.7 K with magnetic field up to 14 T applied perpendicular to the plane. A clear meta-magnetic transition with a small thermal hysteresis is observed between $\mu_0 H_{c1} \sim 5$~T and $\mu_0 H_{c2} \sim 10$~T. In a small field, the magnetization increases linearly with the applied field; beyond $H_{c1}$, the magnetization increases more rapidly with increasing field until it reaches $H_{c2}$. Above this field, the rate of magnetization increase goes back to the original value below $H_{c1}$. The data could be most naturally interpreted by assuming that the ordered moment has both in-plane and out-of-plane component. The in-plane component follows linear M vs. H behavior without any anomaly, while the out-of-plane component goes through a spin-flop transition around 5~T. It appears that the out-of-plane component quickly reaches saturation around $\mu_0 H_{c2} \sim 10$~T. The spin-flop transition of the out-of-plane component can be further analyzed by subtracting the linear in-plane component contribution shown as a dashed line in the main panel. The out-of-plane component does not contribute to the total magnetization below $H_{c1}$. Above this field, the out-of-plane moment goes through spin-flop transition to be parallel to the applied field, which allows additional contribution to the total magnetization. This is the reason for the rapid increase of the magnetization in $H_{c1}< H < H_{c2}$. Above $H_{c2}$, the out-of-plane magnetization reaches saturation and stays constant. The size of increased moment ($\Delta M$) is again order of $\sim 0.1 \mu_B$, consistent with small ordered moment size observed in our neutron scattering experiment.

Finally, we would like to discuss the broad hump around 10-15~K observed in both susceptibility and specific heat measurements. This feature corresponds to the magnetic cusp observed in earlier powder studies \cite{Fletcher1967,Kobayashi1992}. In addition, it is this broad transition that accounts for the bulk of entropy change as shown in Fig.~\ref{fig1-2}(b). Given the quasi-two-dimensional nature of the crystal, it is quite tempting to associate this broad feature with two-dimensional short range ordering, and the sharp peak at 7~K with three-dimensional ordering. This interpretation is supported by recent numerical studies. In particular, Price and Perkins found two magnetic transitions in their recent quantum Monte Carlo investigation of classical Kitaev-Heisenberg model.\cite{Price2012,Price2013} They found that in a purely two-dimensional system, the Kitaev term reduces the continuous symmetry due to Heisenberg interaction, and is responsible for finite temperature long range order. In addition, they found that the phase transition proceeds in two steps. That is, there exists an intermediate state described by Berezinskii-Kosterlitz-Thouless (BKT) critical behavior at higher temperature before the actual transition occurs. We do not observe such a two-dimensional correlation in our neutron scattering data however, perhaps due to poor statistics. Clearly further neutron scattering investigation of the critical behavior is required to elucidate the finite-temperature phase transition of the Kitaev-Heisenberg model.

In summary, we have carried out comprehensive magnetic, thermodynamic, and neutron scattering studies to elucidate the magnetic phase transitions in single crystal $\alpha$-RuCl$_3$. Around 7~K, we found a signature of a magnetic transition in both magnetic susceptibility and specific heat data. We also found a magnetic superlattice peak at (-0.5, 0, 1) from our neutron diffraction experiment below this transition temperature. No magnetic superlattice peaks were observed at (0.5, 0.5, L) positions. These findings are consistent with zigzag type order in the honeycomb plane in this material. However, the ordered moment size is quite small, and the correlation is short-ranged along the direction perpendicular to the plane due to stacking disorder. A broad hump is also observed around 10-15~K, just above the magnetic transition temperature, which could arise from two-dimensional short-range correlation. Our susceptibility and neutron measurements suggest that Kitaev interaction may be quite significant in this compound, and $\alpha$-RuCl$_3$ is an excellent candidate to study Kitaev physics experimentally.

We would like to acknowledge useful discussions with Hae-Young Kee, Ken Burch, Heungsik Kim, and Luke Sandilands. Research at the University of Toronto was supported by Natural Science and Engineering Research Council of Canada, Canada Foundation for Innovation, Ontario Ministry of Research and Innovation, and Canada Research Chair program. The work at National Institute of Standards and Technology is in part supported by the National Science Foundation under Agreement No. DMR-0944772.


\begin{thebibliography}{45}
\bibitem[{Cp_()}]{Cp_note}
\bibinfo{note}{Since we do not know the Debye temperature, and we are mostly
  concerned with non-monotonic features in specific heat, we used a polynomial
  function to approximate phonon contribution. However, exact form of the
  background will likely affect the high temperature tail of the $\Delta C_p$
  data.}
\end{thebibliography}
\end{document}